\documentclass[prl,twocolumn,showpacs]{revtex4-1}
\usepackage{amssymb,amsmath,bm,epsfig,graphicx,subfigure}

\begin{document}

\title{Zero Modes on
 Zero Angle Grain Boundaries in Graphene}
\author{Madeleine Phillips$^1$ and E. J. Mele$^{1,2}$}
    \email{mele@physics.upenn.edu}
    \affiliation{$^1$Department of Physics and Astronomy,  University of Pennsylvania, Philadelphia PA 19104  \\
    $^2$Department of Physics, Loughborough University LE11 3TU, UK}
\date{\today}

\begin{abstract}
Electronic states confined to zero angle grain boundaries in single layer graphene are analyzed using topological band theoretic arguments. We identify a hidden chiral symmetry which supports symmetry protected zero modes in projected bulk gaps. These branches occupy a finite fraction of the interface-projected Brillouin zone and terminate at bulk gap closures, manifesting topological transitions in the occupied manifolds of the bulk systems that are  joined at an interface. These features are studied by numerical calculations on a tight binding lattice and by analysis of the geometric phases of the bulk ground states.

.
\end{abstract}

\pacs{73.20.-r,73.22.Pr,73.20.Hb,3.65.Vf} \maketitle

\medskip
The development of practical methods for the synthesis of large area  single- and few-layer graphenes  \cite{kong,iijima,gbwire,zettl} is focusing attention on the influence of grain boundaries on their electronic behavior  \cite{zettl,muller,kim,gbwire,lambin}. These extended defects have been studied theoretically to understand their reconstruction of the low energy Dirac spectra and their signatures in transport \cite{lambin,LdS1,LdS2,sun,yazyev1,yazyev2,zaanen,gunlycke}.
In this Letter we consider the family of ``zero angle" grain boundaries (ZGB's) and study their electronic properties using quantum geometric methods. In ZGB's the orientation of the crystalline axes is unchanged across an interface but the lattice structure is shifted in phase (Fig. 1). This can potentially produce a topological mismatch between the ground states of its bounding states and localize interfacial modes. Indeed we find that  ZGB's can host topologically protected zero modes in the form of flat bands at the middle of their projected bandgaps, but the conditions supporting this depend nontrivially on the symmetries of the bulk Hamiltonians and the translational symmetries that are preserved at the interface. Gap closures as a function of the parallel momentum $k_\|$ mark bulk topological transitions manifest by critical points where bands of zero modes at the interface emerge from and disappear into the bulk.  Here we develop a theory for these states using the geometric phase for a composite ground state manifold containing $N$ coupled bands.  Our formulation, developed here for the prototypical case of ZGB's, can be generalized to a broad class of twin boundaries between crystalline materials with misoriented symmetry axes.

Figure 1 shows the lattice structures of two prototypical zero angle grain boundaries examined in this work. Panel (a) is the ``5-5" structure, where two honeycomb lattices are joined on a boundary containing a line of pentagon pairs. In panel (b) the related ``5-5-8" structure doubles the period by the insertion of eight-membered rings. The insets show the primitive cells of the bulk structures defined so that the bulk translation vectors are the same on both side of the interface but with the sublattice labels switched.
\begin{figure}
\includegraphics[angle=0,width=\columnwidth]{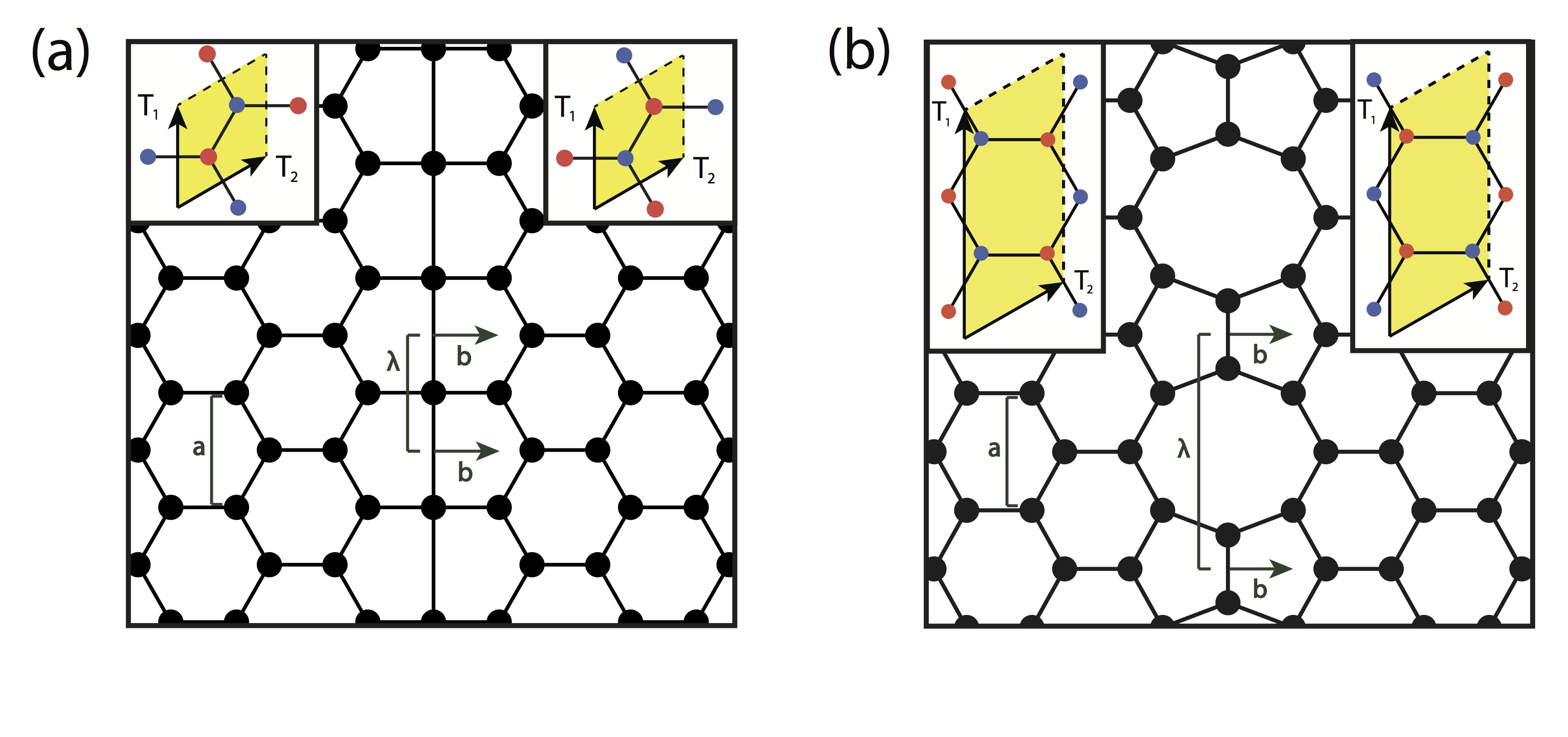}
  \caption{\label{jpeglattices}Lattice structures for the ``55" and ``558" grain boundaries illustrating the Burgers vector $b$ and the period of the Burgers vector lattice $\lambda$. The insets show primitive cells and primitive translation vectors that define the bounding phases to the left
  and right sides of the interface. }
\end{figure}

At first, the prospects for finding zero mode physics in the lattice structures of Figure 1 appear remote. All zero angle grain boundaries contain a macroscopic fraction of odd-membered rings, explicitly breaking the global chiral symmetry of the Hamiltonian which usually is exploited to identify candidate $E=0$ eigenstates. However, this argument is flawed. Both structures in Figure 1 (and its longer period variants) retain a
$x \rightarrow -x$  mirror symmetry and the global Hamiltonian can be partitioned into invariant subspaces that are respectively even and odd under this reflection. Mirror-odd states have a nodal line along the vertical bonds that define its odd membered rings. Removal of these bonds leaves the remaining system bipartite, retaining a chiral symmetry in its projected mirror odd subspace. Formally, one can define a chiral operator ${\cal S}$ that anticommutes with the global Hamiltonian ${\cal H}$.  When written in terms of the original unsymmetrized lattice degrees of freedom, one finds that  ${\cal S}$ is highly nonlocal and contains long range off diagonal amplitudes  that enforce the sign change of its mirror-odd amplitudes on the left and right hand sides of the interface \cite{mpejm}.
\begin{figure}
\includegraphics[angle=0,width=\columnwidth]{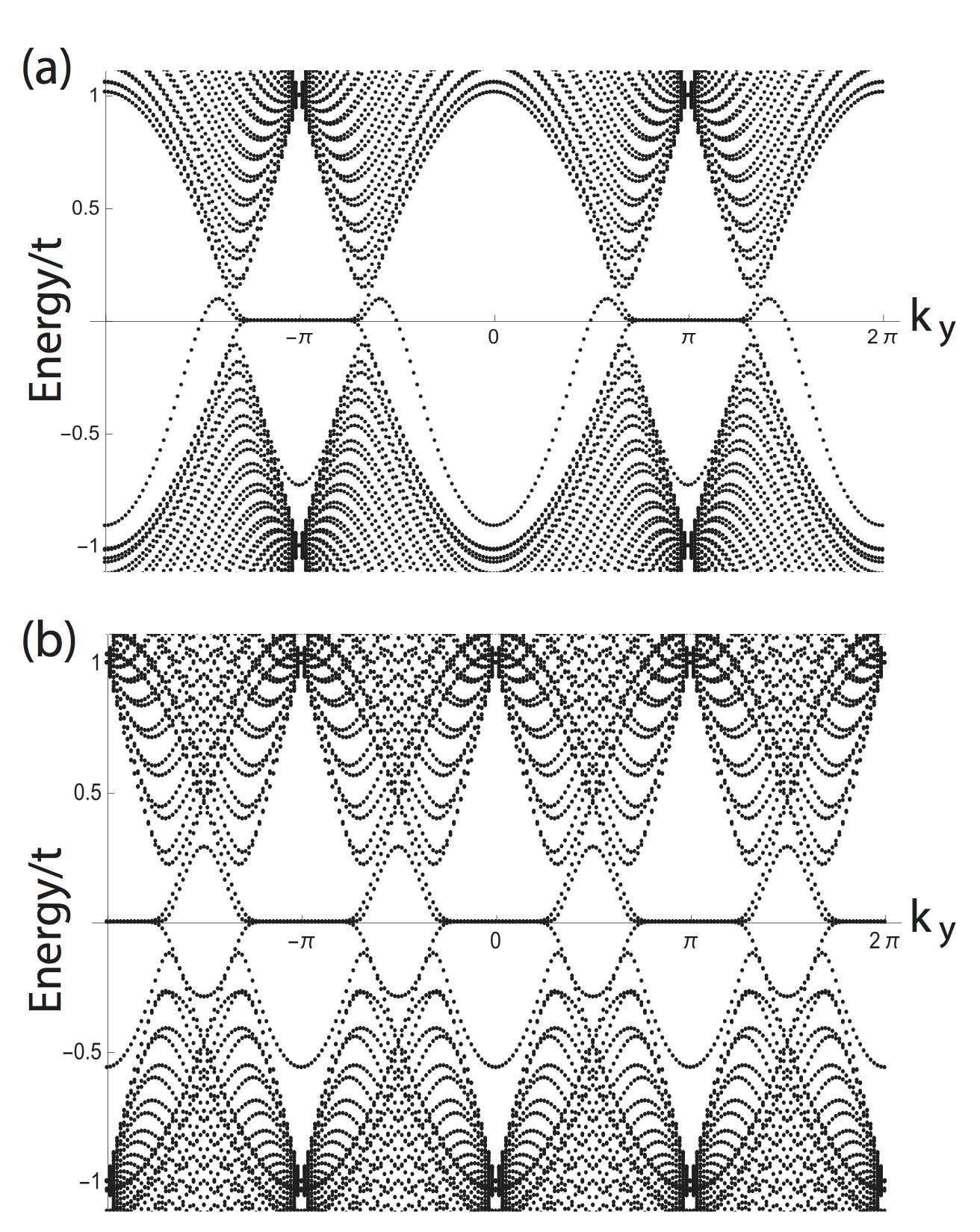}
  \caption{\label{spectra}Tight binding bandstructures for twinned graphene ribbons containing a ``55" grain boundary (top) and a ``558" grain boundary (bottom). Half the projected bandgaps support zero modes confined at the grain boundary, and additional low energy features that are finite size effects associated with edge states on the outer edges of the ribbon.}
\end{figure}
Indeed tight binding lattice calculations carried out on finite width twinned ribbons (Figure 2) show clear evidence of zero mode physics for these structures. One observes that {\it half} the projected gaps host flat bands at $E=0$. These solutions are localized in the grain boundary and reside in the mirror odd subspace.  Note also that near $E=0$ we find additional low energy features that are finite size effects: they are the edge modes localized at the outer zigzag edges of our model.

The lattice structures shown in Figure 1 retain a discrete translational symmetry along the grain boundary,
and the Hamiltonian can be written in Bloch form ${\cal H}(k_y)$.  Each Bloch Hamiltonian presents a one-dimensional problem in the perpendicular ($x$) coordinate with a gapped spectrum except at isolated critical points where the bulk Dirac points project onto the interface. It is tempting  to associate the zero modes with the solution of a Jackiw Rebbi problem induced by a sign reversal of a Dirac mass crossing the grain boundary \cite{jackiw}.

Implemented in its simplest form this interpretation is problematic. For example, note the alternation of projected gaps that do and do not support zero modes in Figure 2. For the ``55" structure the gap that is analytically connected to $k_y=0$ is topologically inactive: there are no zero modes and apparently no sign reversal of a Dirac mass  at the interface. As $k_y$ crosses $2 \pi/3$ this changes and the system supports a zero mode until the next gap closure.  For the related ``558" structure the situation is exactly reversed. Here the projected gaps that are analytically connected to the $k_y=0$ state {\it do} support a zero mode and the smaller gaps (e.g. for $\pi/3 < k_y < 2 \pi/3$) are inactive. This reversal presents a dilemma to a topological interpretation which would identify the zero mode count with a change of the ground state topology of the two bounding states {\it far from the interface}. Inspection of Figure 1 shows that the asymptotic structures of the ``55" and ``558" ZGB's are actually identical. Furthermore, interpreting a putative sign change of a Dirac mass even in the ``active" gaps is subtle. The momentum resolved  Hamiltonians ${\cal H}_{L(R)} (k_y)$ for the bulk states left and right of the interface are unitarily equivalent in {\it every} gapped sector. Physically this reflects the fact that their lattice structures are locally indistinguishable.  Thus one is faced with the problem of identifying precisely ``what is changing" at an interface that produces a mass inversion between two otherwise structurally indistinguishable systems and in a selected subset of their momentum-resolved bandgaps.

 We address these questions by developing a theory of the geometric phase for a twinned bicrystal. To this end it is useful to note that the grain boundary is a periodic array of dislocations. The ZGB's of Figure 1 are linear arrays of {\it partial} dislocations, each with Burgers vector $\textbf{b}=a/\sqrt{3} \, \hat e_x$  inducing a sublattice exchange. The two structures of Figure 1 are distinguished only by the periods of their dislocation lattices: $\lambda = a(2a)$ for the ``55"(``558") ZGB's respectively. In the far field surrounding a single dislocation one can locally define a Bloch Hamiltonian ${\cal H}(\textbf{k})$. Parallel transport of a Bloch state with momentum $\textbf{k}$ in a counterclockwise loop encircling the core accumulates a geometric phase $\theta = \textbf{k} \cdot \textbf{b}$, associated with a $\textbf{k}$-dependent point flux. For the grain boundaries in Figure 1 we choose a linear gauge and represent the topological phase as the phase accumulated in a $\textbf{k}$-dependent effective vector potential
\begin{eqnarray}
{\bf{\cal A}} =  \frac{\textbf{k} \cdot \textbf{b}}{\lambda} \, \Theta(x) \, \hat e_y
\end{eqnarray}
where $\Theta(x)$ is the one-sided step function  $(1 + {\rm sign}(x))/2$.  By virtue of its $\textbf{k}$ dependence ${\bf{\cal A}}$ preserves time reversal symmetry and our choice of gauge preserves
translational symmetry parallel to the interface. Note that ${\bf{\cal A}}$ is curl free except on the grain boundary where it represents a flux sheet that separates the left and right regions. The periodicity of the bulk Hamiltonian in $\textbf{k}$ space is preserved by defining the crystal momentum in Eqn. 1 mod $\textbf{G}$ restricting it to the first Brillouin zone.

The dynamical momentum ${\boldsymbol{\kappa}} = \textbf{k} - {\bf{\cal A}}$ and continuity of the wavefunctions on the boundary equates the (conserved) kinematical momentum $k_y$ on the two sides. The ground states of the  bounding states can be compared by studying the mismatch of the dynamical momentum $\boldsymbol{\kappa}$  induced by ${\bf{\cal A}}$ for a fixed value of $k_y$. As shown in Figure 3, a Brillouin zone tour $\alpha' \rightarrow \beta'$ on the left (ungauged) side at a fixed value of $k_y$ maps to the sloped trajectory $\alpha \rightarrow \beta$ on the right (gauged) side. Note that while the former path is a ``closed" trajectory in $\textbf{k}$ space, i.e. $\textbf{k}_{\alpha'} = \textbf{k}_{\beta'}$, the latter is open, terminating at $\textbf{k}_\alpha$ and $\textbf{k}_\beta$. This momentum offset is an unavoidable consequence of the topological structure of the boundary.  The left and right regions can be compared by studying the evolution of the ground states of a family of Hamiltonians ${\cal H}(\gamma, k_x; k_y)$  (holding $k_y$  constant) and using a parameter $\gamma=0(1)$ to define the ungauged(gauged) sides of the system. We study a closed  $(\gamma,k_x$)-space tour which is a parametric circuit that links the gauged segment $\alpha \rightarrow \beta$ with a return path $\beta' \rightarrow \alpha'$ as shown in Figure 3(lower panel). The geometric phase evaluated along this circuit is gauge invariant and quantifies the difference between the ground states of the two bounding systems. We note that since the bounding Hamiltonians ${\cal H}_{L(R)}(\textbf{k})$ are gauge equivalent we can regard this circuit as a closed momentum space tour in the space of ground states of a {\it single} Hamiltonian.
\begin{figure}
\includegraphics[angle=0,width=\columnwidth]{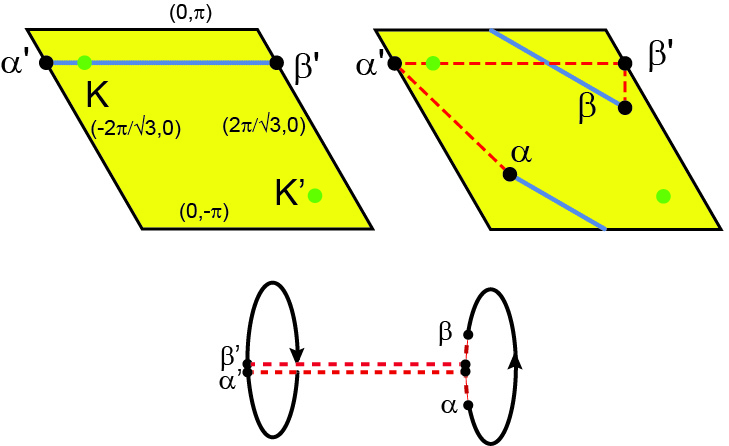}
\caption{\label{tours}Momentum space tours through the Brillouin zone for (left) the ``55" grain boundary at $k_y=2 \pi/3$ for an ungauged trajectory $\alpha' \rightarrow \beta'$  and (right) for its gauged transformed image $\alpha \rightarrow \beta$. The two are linked to form a closed reciprocal space tour through the dashed segments. The green circles are the projections of the $K$ and $K'$ points into the first Brillouin zone. The ungauged segment is critical and passes through a gap closure at a Dirac point. The lower panel illustrates the closed loop that combines these segments to calculate the change of the ground state geometric phase across the boundary for each value of $k_y$.  }
\end{figure}

Figure 4 shows the dispersion of the electronic bands along a closed tour and clearly illustrates the {\it raison d'$ \hat e$tre} for its interfacial zero modes. The top panel is for the ``55" structure which has a primitive bulk unit cell and one occupied spin degenerate band. The critical trajectory shown has an ungauged (flat) segment intersecting a gap closure at a Dirac point. The sloped $\alpha \rightarrow \beta$ segment encounters no such degeneracy at this value of $k_y$. Shifting the fiducial value of $k_y$ allows this system to undergo a band inversion {\it on one side} of the grain boundary, i.e. the $\beta' \rightarrow \alpha'$ return path can develop a band inversion with respect to its gauged $\alpha \rightarrow \beta$ image.

\begin{figure}
\includegraphics[angle=0,width=\columnwidth]{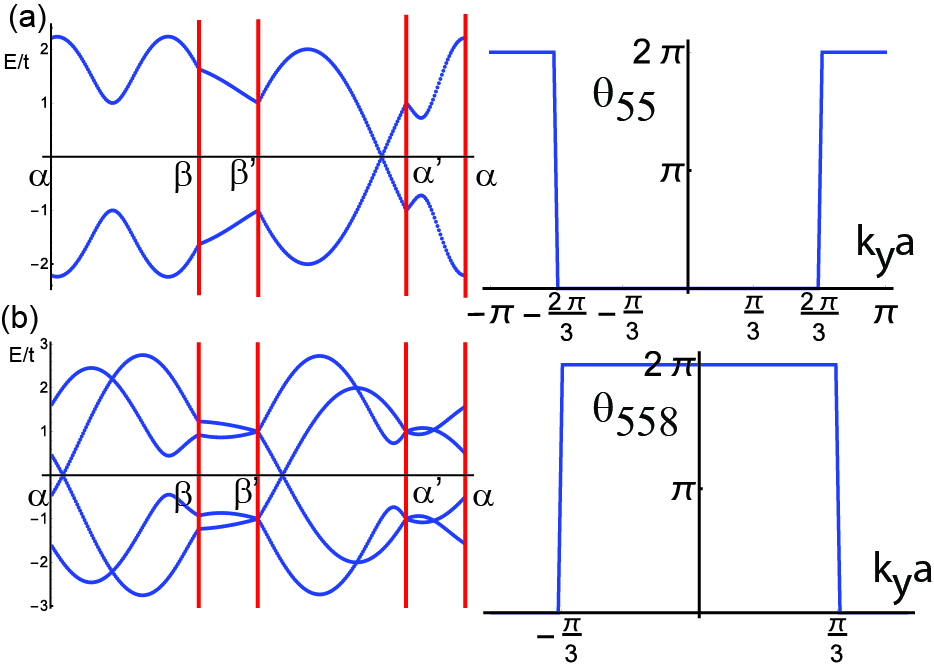}
  \caption{\label{dispersions}Dispersion of electronic bands along a closed momentum space tour (a) for the ``55" grain boundary at $k_y=2 \pi/3$  and (b) for the ``558" grain boundary at $k_y=\pi/3$ (b). In (a) the $\beta' \alpha'$ segment is critical and passes through a Dirac point. In (b) both the ungauged $\beta' \alpha'$ and gauged $\alpha \beta$ segments are critical at a common value of $k_y$ but at opposite Dirac points. The right hand panels show the geometrical phase $\theta$  evaluated along the related paths as a function of $k_y$ revealing topological transitions as $2 \pi$ jump discontinuities.}
\end{figure}

We quantify these observations by calculating the loop integral of the one-band Berry's connection
\begin{eqnarray}
\theta = {\rm Im} \oint \,\, d \boldsymbol{\kappa}  \cdot \langle  u_{\boldsymbol{\kappa}} |\nabla_{\boldsymbol{\kappa}}  u_{\boldsymbol{\kappa}} \rangle
\end{eqnarray}
where $u_{\boldsymbol{\kappa}}$ is the lattice periodic factor of the Bloch wavefunction. For the ``bridge" segments $\alpha' \rightarrow \alpha$ and $\beta \rightarrow \beta'$ we replace the line integral in Eqn. 2 by the finite difference counterparts $\theta ( \alpha' \rightarrow \alpha) = \arg \langle u_{\textbf{k}_\alpha} | u_{\textbf{k}_{\alpha'}} \rangle$ and $\theta ( \beta \rightarrow \beta') = \arg \langle u_{\textbf{k}_{\beta'}} | u_{\textbf{k}_\beta} \rangle $.
The accumulated phase plotted in Figure 4(a) shows jump discontinuities in $\theta_{55}$ at gap closures at $k_y =\pm 2 \pi/3$ from zero in the inactive gaps (left and right states with the same winding number) to $2 \pi$  in its ``topological" gaps (different). We find that the contribution to the accumulated phase from each side (i.e. $\theta_{\beta' \alpha'}$ and $\theta_{\alpha' \alpha \beta \beta'}$) are equal for every value of $k_y$.  This is physically understandable since the left-right asymmetry of our $\textbf{k}$-space construction is a gauge choice and the interfacial modes are ultimately a property of the joined system.

The interfacial zero mode for the ``55" grain boundary can be understood in terms of the well known midgap surface state that appears at the termination of a graphene sheet on a zigzag edge \cite{nakada}. The $x=0$ nodal line in the mirror-odd subspace is a hard wall boundary condition that effectively disconnects two halves of a graphene sheet exposing zigzag edges. This supports {\it two} branches of edge states in the small projected gaps arising from the left and right hand surfaces. The mirror-even combination of these states is gapped out by coupling across the grain boundary, while the mirror-odd combination survives as a localized midgap state.

The related ``558" grain boundary structure doubles the lattice period along the grain boundary.  The essential complication in this nonprimitive situation is illustrated in Figure 4(b). The folding of the Brillouin zone produces two occupied (orbital) branches which are required to degenerate at the endpoints $\alpha'$ and $\beta'$.  The occupied manifold therefore unavoidably entangles these two orbital degrees of freedom and its geometric phase must remain invariant under $\textbf{k}$-dependent unitary rotations in the occupied subspace.  This is not a property of sums of the individual band-projected Berry's phases but it can be understood using the matrix-valued connection
\begin{eqnarray}
\chi_{m,n}(\textbf{k},\textbf{k} + \delta \textbf{k}) = \langle u_ {m,\textbf{k}+ \delta \textbf{k}} | u_{n,\textbf{k} } \rangle
\end{eqnarray}
where $m$ and $n$ are occupied bands and $\delta \textbf{k}$ is along the tangent line of the tour. The geometric phase in Eqn. 2 generalizes to the accumulated phase of the loop product of $2 \times 2$ matrices $\boldsymbol\chi(\textbf{k},\textbf{k} + \delta \textbf{k})$ over the the relevant closed tour in $\textbf{k}$ space
\begin{eqnarray}
\theta_{558} = \arg \prod_{\textbf{k}}  \det \boldsymbol\chi(\textbf{k},\textbf{k} + \delta \textbf{k})
\end{eqnarray}
Similar to the one-band case, the contributions from the bridging links $\alpha' \rightarrow \alpha$ and $\beta \rightarrow \beta'$ are given by finite difference expressions: $
 \theta (\alpha' \rightarrow \alpha) = \arg \det \boldsymbol\chi(\textbf{k}_\alpha',\boldsymbol\kappa_{\alpha})$ and $\theta (\beta \rightarrow \beta') = \arg \det  \boldsymbol\chi(\boldsymbol\kappa_{\beta},\textbf{k}_{\beta'}) $. Interestingly, we find that for the ``558" grain boundary the $\alpha \rightarrow \beta$ and $\beta' \rightarrow \alpha'$ segments each undergo simultaneous gap closures at $k_y=\pi/3$  though these occur at the {\it opposite} projected bulk Dirac points. This also signals a relative inversion of the bulk bands as revealed in the total accumulated phase plotted as a function of $k_y$ in Fig. 4(b). This identifies the larger gaps centered on the origin ($-\pi/3 < k_y < \pi/3$) as topologically mismatched and the smaller gaps ($\pi/3 < k_y < 2 \pi/3$) as silent, in agreement with the numerical results of Figure 2.

 Our construction demonstrates that although the far field lattices are identical for the ``55" and ``558" grain boundaries, they retain information about the translational symmetry that is present on the boundary. This is encoded in the degeneracies and symmetries of the  bulk eigenstates (a bulk property) folded into their reconstructed reduced zones (an interface property).  This information is intrinsically nonlocal and is neatly quantified by consideration of the geometric phase evaluated over a momentum space loop that bridges the two half spaces in the appropriate composite $N$-band manifold.

  For both grain boundaries the bulk Hamiltonians at a fixed value of the projected momentum $k_y$ are members of the Altland Zirnbauer chiral unitary class (AIII) \cite{ludwig,altland}.  These Hamiltonians preserve chiral (sublattice) symmetry but they have broken time reversal symmetry (here explicitly broken for a generic value of $k_y$) and broken charge conjugation symmetry.  In $d=1$ the ground states in this class are topologically nontrivial and are distinguished by an integer-valued topological index.  The winding numbers calculated above measure the interfacial mismatch of this index as a function of $k_y$, thereby counting the number of zero modes (flat bands) in each projected gap.

Our approach admits a straightforward generalization to twin boundaries with nonzero rotation angle and to grain boundaries that embed full dislocations
(Burgers vector is a lattice translation rather than a fractional translation).  Our preliminary work indicates that this distinction is important and can control the topological character of the boundary \cite{mpejm}. It will also be useful to augment this topological analysis to address the consequences of local symmetry breaking perturbations (presumed to be weak) that inevitably occur in atomistic models that suggest structure-specific spectral reconstruction near the neutrality point \cite{lambin,LdS1,LdS2,sun,yazyev1,yazyev2,zaanen,gunlycke}. Finally the presence of flat or weakly dispersing bands near charge neutrality invites an investigation of its interaction-induced instabilities.

This work is supported by the Department of Energy under grant ER45118. EJM acknowledges the generous support from the Leverhulme Trust while visiting Loughborough University where this work was completed.

\end{document}